# Comment on "Prediction of lattice constant in cubic perovskites"

Roberto L. Moreira[a,*] and Anderson Dias[b]

[a]*Departamento de Física, ICEx,UFMG, C.P. 702, Belo Horizonte-MG, 30123-970, Brazil*

[b]*Departamento de Química, ICEB, UFOP, Ouro Preto-MG, 35400-000, Brazil*

**Abstract**

In a recent work by Jiang *et al*. [1], the interrelationship between lattice constant, ionic radii and tolerance factor of cubic perovskites has been established and an empirical equation was obtained. However, the assumption of incorrect ionic coordination led to an incorrect mathematical expression even though the average relative errors between predicted and observed lattice constants of 132 materials were below 1%. Here, corrected coefficients for that empirical expression are obtained, which would likely be useful for investigation of general perovskite materials.

[*]Corresponding author. Tel.: +55-31-3499-5624

*E-mail address*: bmoreira@fisica.ufmg.br

The prediction of lattice constant values for perovskite materials is of recognized importance, owing to the development of new materials designed for different applications, such as ferroelectric thin films, microwave and semiconductor technologies *etc* [2-6]. Therefore, the methodology developed by Jiang *et al.* [1], which allows one to predict the lattice constants ($a_{pred}$) of cubic perovskites ($ABX_3$) by using the known ionic radii of the cations and anion, appears to be very useful and should likely become a key reference for materials scientists and engineers working in the field [7-10]. However, in their work, Jiang *et al.* used ionic radii of all ions *A*, *B* and *X*, for coordination numbers (CN) equal to 6, obtaining an incorrect expression for $a_{pred}$ as a function of the ionic radii and tolerance factor. Once we truly believe on the importance of that procedure, the purpose of this work is to present the corrected coefficients for the Jiang's expression.

In perovskite structures, *B* cations are coordinated by six *X* anions, while *A* cations present CN = 12 (also coordinated by *X* anions). The *X* anions have CN = 2, being coordinated by two *A* cations, since the distance *A*-O is about 40% larger than the *B*-O bond distance. The correct ionic radii ($r_A$, $r_B$, $r_X$), taken from Shannon's work [11] and other references [12-16] are presented in Table 1. Tolerance factors $\{t = (r_A + r_X) / [\sqrt{2}\,(r_B + r_X)]\}$ were recalculated and new coefficients (β, δ and γ) for the Jiang's expression, $a_{pred} = 2\beta\,(r_B + r_X) + \gamma t - \delta$, were obtained. These coefficients are summarized in Tables 1 and 2, together with two sets of Jiang's coefficients and with the average relative errors for the 132 materials used for determining the Jiang's expression. The second set (Jiang[b]) was determined after using the correct β value obtained by those authors.

[TABLES 1 AND 2]

The difference in the empirical coefficients can be easily seen from the plots of lattice constant (*a*) *versus* $2(r_B + r_X)$, presented in Fig. 1, and $[2\beta(r_B + r_X) - a]$ *versus* the tolerance factor (*t*), Fig. 2, with the $\beta$ values determined from the fits of Fig. 1. It can be seen that the correction of the ionic radii of *A* and *X* ions led to smaller $2(r_B + r_X)$ and larger *t* values, shifting the data in the figures and changing the fitting parameters. However, the linear trends in all cases are maintained, and the fits of the curves give $a = \alpha + 2\beta\ (r_B + r_X)$ in Fig. 1, and $2\beta\ (r_B + r_X) - a = -\gamma t + \delta$, in Fig. 2.

[FIGURES 1 AND 2]

It is worthy noticing that an ideal cubic perovskite would have $t \approx 1$, while materials with smaller *t* values usually belong to lower symmetry structures, with tilted $BX_6$ octahedra [17]. The critical *t* value for stabilizing cubic structures is about 0.985 [17]. This value is higher than those obtained in ref. [1] for all the 132 materials investigated. Take for instance $SrTiO_3$: according to ref. [1], this material would present $t = 0.908$, indicating that it would belong to a distorted/tilted non-cubic structure; after the correction proposed here, *t* becomes 1.009, in agreement with its experimentally observed untilted cubic structure (at room temperature) [18]. Another remark concerns the increase in the average deviation between predicted and experimental lattice parameters. As it can be seen in Fig. 2, the dispersion of data increased after correction of the CN's. This could be due to a lower accuracy of ionic radii used for CN=12 (worse statistics than for CN=6), although we should remember that most structures are indeed pseudocubic, with tilted $BX_6$ octahedra, which change both, the approximated *a* value (i.e., the cubic ratio of the unit cell volume) as well as the "average" ionic radii used in the calculations.

In conclusion, we believe that the empirical expression for determining the lattice parameters of cubic perovskites developed in ref. [1], with the corrections presented here, could be very useful in future researches in this field.

**Acknowledgements**

This work was partially supported by the Brazilian agencies MCT/CNPq, FINEP and FAPEMIG.

Table 1 – Ionic radii and lattice parameters for 132 perovskites (compound 57 is meaningless). Ionic radii for Cl, Br, I, Ag, Sn, and lanthanides from refs. 12-16.

| No. | Compound | $a$ (Å) | $r_A$ (Å) | $r_B$ (Å) | $r_X$ (Å) | $t$ | $2(r_B + r_X)$ | $2\beta(r_B + r_X) - a$ | $a_{pred}$ (Å) | dev (%) |
|---|---|---|---|---|---|---|---|---|---|---|
| 1 | $CsIO_3$ | 4.674 | 1.88 | 0.95 | 1.35 | 0.993 | 4.60 | -0.484 | 4.539 | 2.883 |
| 2 | $RbUO_3$ | 4.323 | 1.72 | 0.76 | 1.35 | 1.029 | 4.22 | -0.479 | 4.234 | 2.064 |
| 3 | $KUO_3$ | 4.29 | 1.64 | 0.76 | 1.35 | 1.002 | 4.22 | -0.446 | 4.203 | 2.021 |
| 4 | $RbPaO_3$ | 4.368 | 1.72 | 0.78 | 1.35 | 1.019 | 4.26 | -0.488 | 4.259 | 2.490 |
| 5 | $KPaO_3$ | 4.341 | 1.64 | 0.78 | 1.35 | 0.993 | 4.26 | -0.461 | 4.229 | 2.579 |
| 6 | $KTaO_3$ | 3.988 | 1.64 | 0.64 | 1.35 | 1.062 | 3.98 | -0.363 | 4.053 | 1.638 |
| 7 | $BaFeO_3$ | 3.994 | 1.61 | 0.585 | 1.35 | 1.082 | 3.87 | -0.469 | 3.975 | 0.476 |
| 8 | $BaMoO_3$ | 4.040 | 1.61 | 0.65 | 1.35 | 1.047 | 4.00 | -0.396 | 4.053 | 0.333 |
| 9 | $BaNbO_3$ | 4.080 | 1.61 | 0.68 | 1.35 | 1.031 | 4.06 | -0.382 | 4.091 | 0.259 |
| 10 | $BaSnO_3$ | 4.116 | 1.61 | 0.69 | 1.35 | 1.026 | 4.08 | -0.400 | 4.103 | 0.315 |
| 11 | $BaHfO_3$ | 4.171 | 1.61 | 0.71 | 1.35 | 1.016 | 4.12 | -0.418 | 4.128 | 1.027 |
| 12 | $BaZrO_3$ | 4.193 | 1.61 | 0.72 | 1.35 | 1.011 | 4.14 | -0.422 | 4.141 | 1.245 |
| 13 | $BaIrO_3$ | 4.100 | 1.61 | 0.625 | 1.35 | 1.060 | 3.95 | -0.502 | 4.023 | 1.879 |
| 14 | $EuTiO_3$ | 3.905 | 1.23 | 0.67 | 1.35 | 0.903 | 4.04 | -0.225 | 3.927 | 0.565 |
| 15 | $NaWO_3$ | 3.850 | 1.39 | 0.62 | 1.35 | 0.983 | 3.94 | -0.261 | 3.927 | 2.006 |
| 16 | $SnTaO_3$ | 3.880 | 1.10 | 0.68 | 1.35 | 0.853 | 4.06 | -0.182 | 3.889 | 0.227 |
| 17 | $SrMnO_3$ | 3.806 | 1.44 | 0.53 | 1.35 | 1.049 | 3.76 | -0.381 | 3.838 | 0.843 |
| 18 | $SrVO_3$ | 3.890 | 1.44 | 0.58 | 1.35 | 1.022 | 3.86 | -0.374 | 3.898 | 0.214 |
| 19 | $SrFeO_3$ | 3.850 | 1.44 | 0.585 | 1.35 | 1.020 | 3.87 | -0.325 | 3.904 | 1.415 |
| 20 | $SrTiO_3$ | 3.905 | 1.44 | 0.605 | 1.35 | 1.009 | 3.91 | -0.343 | 3.929 | 0.615 |
| 21 | $SrTcO_3$ | 3.949 | 1.44 | 0.645 | 1.35 | 0.989 | 3.99 | -0.315 | 3.979 | 0.757 |
| 22 | $SrMoO_3$ | 3.975 | 1.44 | 0.65 | 1.35 | 0.986 | 4.00 | -0.331 | 3.985 | 0.257 |
| 23 | $SrNbO_3$ | 4.016 | 1.44 | 0.68 | 1.35 | 0.972 | 4.06 | -0.318 | 4.023 | 0.182 |
| 24 | $SrSnO_3$ | 4.034 | 1.44 | 0.69 | 1.35 | 0.967 | 4.08 | -0.318 | 4.036 | 0.052 |
| 25 | $SrHfO_3$ | 4.069 | 1.44 | 0.71 | 1.35 | 0.958 | 4.12 | -0.316 | 4.062 | 0.175 |
| 26 | $CaVO_3$ | 3.767 | 1.34 | 0.58 | 1.35 | 0.986 | 3.86 | -0.251 | 3.857 | 2.381 |
| 27 | $BaPbO_3$ | 4.265 | 1.61 | 0.775 | 1.35 | 0.985 | 4.25 | -0.394 | 4.211 | 1.260 |
| 28 | $BaTbO_3$ | 4.285 | 1.61 | 0.76 | 1.35 | 0.992 | 4.22 | -0.441 | 4.192 | 2.173 |
| 29 | $BaPrO_3$ | 4.354 | 1.61 | 0.85 | 1.35 | 0.951 | 4.40 | -0.346 | 4.310 | 1.016 |
| 30 | $BaCeO_3$ | 4.397 | 1.61 | 0.87 | 1.35 | 0.943 | 4.44 | -0.353 | 4.336 | 1.376 |
| 31 | $BaAmO_3$ | 4.357 | 1.61 | 0.85 | 1.35 | 0.951 | 4.40 | -0.349 | 4.310 | 1.084 |
| 32 | $BaNpO_3$ | 4.384 | 1.61 | 0.87 | 1.35 | 0.943 | 4.44 | -0.340 | 4.336 | 1.084 |
| 33 | $BaUO_3$ | 4.387 | 1.61 | 0.89 | 1.35 | 0.934 | 4.48 | -0.306 | 4.363 | 0.539 |
| 34 | $BaPaO_3$ | 4.450 | 1.61 | 0.9 | 1.35 | 0.930 | 4.50 | -0.351 | 4.377 | 1.644 |
| 35 | $BaThO_3$ | 4.480 | 1.61 | 0.94 | 1.35 | 0.914 | 4.58 | -0.308 | 4.431 | 1.088 |
| 36 | $SrTbO_3$ | 4.180 | 1.44 | 0.76 | 1.35 | 0.935 | 4.22 | -0.336 | 4.127 | 1.263 |
| 37 | $SrAmO_3$ | 4.230 | 1.44 | 0.85 | 1.35 | 0.897 | 4.40 | -0.222 | 4.248 | 0.419 |
| 38 | $SrPuO_3$ | 4.280 | 1.44 | 0.86 | 1.35 | 0.893 | 4.42 | -0.254 | 4.261 | 0.436 |
| 39 | $SrCoO_3$ | 3.850 | 1.44 | 0.53 | 1.35 | 1.049 | 3.76 | -0.425 | 3.838 | 0.309 |
| 40 | $BaTiO_3$ | 4.012 | 1.61 | 0.605 | 1.35 | 1.071 | 3.91 | -0.450 | 3.999 | 0.328 |
| 41 | $CaTiO_3$ | 3.840 | 1.34 | 0.605 | 1.35 | 0.973 | 3.91 | -0.278 | 3.888 | 1.248 |
| 42 | $CeAlO_3$ | 3.772 | 1.34 | 0.535 | 1.35 | 1.009 | 3.77 | -0.338 | 3.801 | 0.781 |
| 43 | $EuAlO_3$ | 3.725 | 1.23 | 0.535 | 1.35 | 0.968 | 3.77 | -0.291 | 3.755 | 0.794 |
| 44 | $EuCrO_3$ | 3.803 | 1.23 | 0.615 | 1.35 | 0.928 | 3.93 | -0.223 | 3.856 | 1.383 |
| 45 | $EuFeO_3$ | 3.836 | 1.23 | 0.645 | 1.35 | 0.914 | 3.99 | -0.206 | 3.894 | 1.522 |
| 46 | $GdAlO_3$ | 3.710 | 1.22 | 0.535 | 1.35 | 0.964 | 3.77 | -0.276 | 3.750 | 1.087 |
| 47 | $GdCrO_3$ | 3.795 | 1.22 | 0.615 | 1.35 | 0.925 | 3.93 | -0.215 | 3.852 | 1.489 |
| 48 | $GdFeO_3$ | 3.820 | 1.22 | 0.645 | 1.35 | 0.911 | 3.99 | -0.186 | 3.890 | 1.842 |
| 49 | $KNbO_3$ | 4.007 | 1.64 | 0.64 | 1.35 | 1.062 | 3.98 | -0.382 | 4.053 | 1.156 |
| 50 | $LaAlO_3$ | 3.778 | 1.36 | 0.535 | 1.35 | 1.017 | 3.77 | -0.344 | 3.810 | 0.846 |
| 51 | $LaCrO_3$ | 3.874 | 1.36 | 0.615 | 1.35 | 0.975 | 3.93 | -0.294 | 3.909 | 0.896 |
| 52 | $LaFeO_3$ | 3.920 | 1.36 | 0.645 | 1.35 | 0.961 | 3.99 | -0.286 | 3.947 | 0.681 |
| 53 | $LaGaO_3$ | 3.874 | 1.36 | 0.62 | 1.35 | 0.973 | 3.94 | -0.285 | 3.915 | 1.058 |
| 54 | $LaRhO_3$ | 3.940 | 1.36 | 0.665 | 1.35 | 0.951 | 4.03 | -0.269 | 3.972 | 0.820 |
| 55 | $LaTiO_3$ | 3.920 | 1.36 | 0.67 | 1.35 | 0.949 | 4.04 | -0.240 | 3.979 | 1.499 |
| 56 | $LaVO_3$ | 3.910 | 1.36 | 0.64 | 1.35 | 0.963 | 3.98 | -0.285 | 3.940 | 0.776 |
| 57 | $NaAlO_3$ | 3.730 | 1.39 | 0.535 | 1.35 | 1.028 | 3.77 | -0.296 | 3.823 | 2.486 |
| 58 | $NaTaO_3$ | 3.881 | 1.39 | 0.64 | 1.35 | 0.974 | 3.98 | -0.256 | 3.952 | 1.841 |
| 59 | $NdAlO_3$ | 3.752 | 1.27 | 0.535 | 1.35 | 0.983 | 3.77 | -0.320 | 3.772 | 0.523 |
| 60 | $NdCoO_3$ | 3.777 | 1.27 | 0.545 | 1.35 | 0.978 | 3.79 | -0.325 | 3.784 | 0.184 |
| 61 | $NdCrO_3$ | 3.835 | 1.27 | 0.615 | 1.35 | 0.943 | 3.93 | -0.255 | 3.872 | 0.963 |
| 62 | $NdFeO_3$ | 3.870 | 1.27 | 0.645 | 1.35 | 0.929 | 3.99 | -0.236 | 3.910 | 1.046 |
| 63 | $NdMnO_3$ | 3.800 | 1.27 | 0.645 | 1.35 | 0.929 | 3.99 | -0.166 | 3.910 | 2.907 |
| 64 | $PrAlO_3$ | 3.757 | 1.30 | 0.535 | 1.35 | 0.994 | 3.77 | -0.323 | 3.784 | 0.729 |
| 65 | $PrCrO_3$ | 3.852 | 1.30 | 0.615 | 1.35 | 0.954 | 3.93 | -0.272 | 3.884 | 0.835 |
| 66 | $PrFeO_3$ | 3.887 | 1.30 | 0.645 | 1.35 | 0.939 | 3.99 | -0.253 | 3.923 | 0.915 |
| 67 | $PrGaO_3$ | 3.863 | 1.30 | 0.62 | 1.35 | 0.951 | 3.94 | -0.274 | 3.891 | 0.713 |
| 68 | $PrMnO_3$ | 3.820 | 1.30 | 0.645 | 1.35 | 0.939 | 3.99 | -0.186 | 3.923 | 2.685 |
| 69 | $PrVO_3$ | 3.890 | 1.30 | 0.64 | 1.35 | 0.942 | 3.98 | -0.265 | 3.916 | 0.672 |
| 70 | $SmAlO_3$ | 3.734 | 1.24 | 0.535 | 1.35 | 0.972 | 3.77 | -0.300 | 3.759 | 0.665 |
| 71 | $SmCoO_3$ | 3.750 | 1.24 | 0.545 | 1.35 | 0.966 | 3.79 | -0.298 | 3.771 | 0.567 |
| 72 | $SmVO_3$ | 3.890 | 1.24 | 0.64 | 1.35 | 0.920 | 3.98 | -0.265 | 3.892 | 0.049 |
| 73 | $SmFeO_3$ | 3.845 | 1.24 | 0.645 | 1.35 | 0.918 | 3.99 | -0.211 | 3.898 | 1.389 |
| 74 | $SrZrO_3$ | 4.101 | 1.44 | 0.72 | 1.35 | 0.953 | 4.14 | -0.330 | 4.075 | 0.638 |
| 75 | $YAlO_3$ | 3.680 | 1.20 | 0.535 | 1.35 | 0.957 | 3.77 | -0.246 | 3.742 | 1.679 |
| 76 | $YCrO_3$ | 3.768 | 1.20 | 0.615 | 1.35 | 0.918 | 3.93 | -0.188 | 3.843 | 1.999 |
| 77 | $YFeO_3$ | 3.785 | 1.20 | 0.645 | 1.35 | 0.904 | 3.99 | -0.151 | 3.882 | 2.570 |
| 78 | $CsCdF_3$ | 4.470 | 1.88 | 0.95 | 1.285 | 1.001 | 4.47 | -0.398 | 4.430 | 0.888 |
| 79 | $CsCaF_3$ | 4.523 | 1.88 | 1.00 | 1.285 | 0.979 | 4.57 | -0.360 | 4.496 | 0.586 |
| 80 | $CsHgF_3$ | 4.570 | 1.88 | 1.02 | 1.285 | 0.971 | 4.61 | -0.371 | 4.523 | 1.023 |
| 81 | $CsSrF_3$ | 4.750 | 1.88 | 1.18 | 1.285 | 0.908 | 4.93 | -0.259 | 4.743 | 0.147 |
| 82 | $TlCoF_3$ | 4.138 | 1.70 | 0.745 | 1.285 | 1.040 | 4.06 | -0.440 | 4.100 | 0.907 |
| 83 | $TlFeF_3$ | 4.188 | 1.70 | 0.78 | 1.285 | 1.022 | 4.13 | -0.426 | 4.144 | 1.046 |
| 84 | $TlMnF_3$ | 4.260 | 1.70 | 0.83 | 1.285 | 0.998 | 4.23 | -0.407 | 4.208 | 1.224 |
| 85 | $TlCdF_3$ | 4.400 | 1.70 | 0.95 | 1.285 | 0.944 | 4.47 | -0.328 | 4.366 | 0.782 |
| 86 | $NH_4ZnF_3$ | 4.115 | 1.80 | 0.74 | 1.285 | 1.077 | 4.05 | -0.426 | 4.134 | 0.459 |
| 87 | $NH_4CoF_3$ | 4.129 | 1.80 | 0.745 | 1.285 | 1.075 | 4.06 | -0.431 | 4.140 | 0.266 |
| 88 | $NH_4FeF_3$ | 4.177 | 1.80 | 0.78 | 1.285 | 1.056 | 4.13 | -0.415 | 4.183 | 0.145 |
| 89 | $NH_4MnF_3$ | 4.241 | 1.80 | 0.83 | 1.285 | 1.031 | 4.23 | -0.388 | 4.246 | 0.113 |
| 90 | $RbZnF_3$ | 4.122 | 1.72 | 0.74 | 1.285 | 1.049 | 4.05 | -0.433 | 4.102 | 0.481 |
| 91 | $RbCoF_3$ | 4.141 | 1.72 | 0.745 | 1.285 | 1.047 | 4.06 | -0.443 | 4.108 | 0.788 |
| 92 | $RbVF_3$ | 4.170 | 1.72 | 0.79 | 1.285 | 1.024 | 4.15 | -0.390 | 4.165 | 0.131 |
| 93 | $RbFeF_3$ | 4.174 | 1.72 | 0.78 | 1.285 | 1.029 | 4.13 | -0.412 | 4.152 | 0.528 |
| 94 | $RbMnF_3$ | 4.240 | 1.72 | 0.83 | 1.285 | 1.005 | 4.23 | -0.387 | 4.215 | 0.579 |
| 95 | $RbCdF_3$ | 4.398 | 1.72 | 0.95 | 1.285 | 0.951 | 4.47 | -0.326 | 4.373 | 0.573 |
| 96 | $RbCaF_3$ | 4.452 | 1.72 | 1.00 | 1.285 | 0.930 | 4.57 | -0.289 | 4.440 | 0.264 |
| 97 | $RbHgF_3$ | 4.470 | 1.72 | 1.02 | 1.285 | 0.922 | 4.61 | -0.271 | 4.468 | 0.055 |
| 98 | $KCdF_3$ | 4.293 | 1.64 | 0.95 | 1.285 | 0.925 | 4.47 | -0.221 | 4.344 | 1.189 |
| 99 | $KMgF_3$ | 3.989 | 1.64 | 0.72 | 1.285 | 1.032 | 4.01 | -0.336 | 4.046 | 1.419 |
| 100 | $KNiF_3$ | 4.013 | 1.64 | 0.69 | 1.285 | 1.047 | 3.95 | -0.415 | 4.009 | 0.106 |
| 101 | $KZnF_3$ | 4.056 | 1.64 | 0.74 | 1.285 | 1.021 | 4.05 | -0.367 | 4.070 | 0.356 |
| 102 | $KCoF_3$ | 4.071 | 1.64 | 0.745 | 1.285 | 1.019 | 4.06 | -0.373 | 4.077 | 0.140 |
| 103 | $KVF_3$ | 4.100 | 1.64 | 0.79 | 1.285 | 0.997 | 4.15 | -0.320 | 4.134 | 0.819 |
| 104 | $KFeF_3$ | 4.121 | 1.64 | 0.78 | 1.285 | 1.002 | 4.13 | -0.359 | 4.121 | 0.003 |
| 105 | $KMnF_3$ | 4.189 | 1.64 | 0.83 | 1.285 | 0.978 | 4.23 | -0.336 | 4.185 | 0.094 |
| 106 | $AgMgF_3$ | 3.918 | 1.48 | 0.72 | 1.285 | 0.975 | 4.01 | -0.265 | 3.982 | 1.621 |
| 107 | $AgNiF_3$ | 3.936 | 1.48 | 0.69 | 1.285 | 0.990 | 3.95 | -0.338 | 3.944 | 0.195 |
| 108 | $AgZnF_3$ | 3.972 | 1.48 | 0.74 | 1.285 | 0.966 | 4.05 | -0.283 | 4.007 | 0.881 |
| 109 | $AgCoF_3$ | 3.983 | 1.48 | 0.745 | 1.285 | 0.963 | 4.06 | -0.285 | 4.013 | 0.764 |
| 110 | $AgMnF_3$ | 4.030 | 1.48 | 0.83 | 1.285 | 0.924 | 4.23 | -0.177 | 4.124 | 2.340 |
| 111 | $NaVF_3$ | 3.940 | 1.39 | 0.79 | 1.285 | 0.912 | 4.15 | -0.160 | 4.037 | 2.458 |
| 112 | $RbPdF_3$ | 4.298 | 1.72 | 0.86 | 1.285 | 0.991 | 4.29 | -0.390 | 4.254 | 1.021 |
| 113 | $NH_4MgF_3$ | 4.060 | 1.80 | 0.72 | 1.285 | 1.088 | 4.01 | -0.407 | 4.110 | 1.223 |
| 114 | $TlPdF_3$ | 4.301 | 1.70 | 0.86 | 1.285 | 0.984 | 4.29 | -0.393 | 4.247 | 1.264 |
| 115 | $LiBaF_3$ | 3.992 | 1.61 | 0.76 | 1.285 | 1.001 | 4.09 | -0.267 | 4.084 | 2.299 |
| 116 | $RbYbF_3$ | 4.530 | 1.72 | 1.02 | 1.285 | 0.922 | 4.61 | -0.331 | 4.468 | 1.379 |
| 117 | $CsEuF_3$ | 4.780 | 1.88 | 1.17 | 1.285 | 0.912 | 4.91 | -0.308 | 4.729 | 1.064 |
| 118 | $CsPbF_3$ | 4.800 | 1.88 | 1.19 | 1.285 | 0.904 | 4.95 | -0.291 | 4.757 | 0.891 |
| 119 | $CsYbF_3$ | 4.610 | 1.88 | 1.02 | 1.285 | 0.971 | 4.61 | -0.411 | 4.523 | 1.882 |
| 120 | $RbPbCl_3$ | 4.790 | 1.72 | 1.19 | 1.285 | 0.859 | 4.95 | -0.281 | 4.705 | 1.768 |
| 121 | $CsCaCl_3$ | 5.396 | 1.88 | 1.00 | 1.79 | 0.930 | 5.58 | -0.313 | 5.360 | 0.658 |
| 122 | $CsCdCl_3$ | 5.210 | 1.88 | 0.95 | 1.79 | 0.947 | 5.48 | -0.218 | 5.289 | 1.510 |
| 123 | $CsPbCl_3$ | 5.605 | 1.88 | 1.19 | 1.79 | 0.871 | 5.96 | -0.176 | 5.639 | 0.612 |
| 124 | $CsHgCl_3$ | 5.410 | 1.88 | 1.02 | 1.79 | 0.924 | 5.62 | -0.291 | 5.389 | 0.381 |
| 125 | $TlMnCl_3$ | 5.020 | 1.70 | 0.83 | 1.79 | 0.942 | 5.24 | -0.247 | 5.064 | 0.880 |
| 126 | $CsEuCl_3$ | 5.627 | 1.88 | 1.17 | 1.79 | 0.877 | 5.92 | -0.235 | 5.610 | 0.311 |
| 127 | $CsTmCl_3$ | 5.476 | 1.88 | 1.03 | 1.79 | 0.920 | 5.64 | -0.339 | 5.404 | 1.317 |
| 128 | $CsYbCl_3$ | 5.437 | 1.88 | 1.02 | 1.79 | 0.924 | 5.62 | -0.318 | 5.389 | 0.875 |
| 129 | $CsHgBr_3$ | 5.770 | 1.88 | 1.02 | 1.95 | 0.912 | 5.94 | -0.359 | 5.668 | 1.774 |
| 130 | $CsPbBr_3$ | 5.874 | 1.88 | 1.19 | 1.95 | 0.862 | 6.28 | -0.154 | 5.921 | 0.805 |
| 131 | $CsSnBr_3$ | 5.795 | 1.88 | 0.95 | 1.95 | 0.934 | 5.80 | -0.512 | 5.565 | 3.967 |
| 132 | $CsSnI_3$ | 6.219 | 1.88 | 0.95 | 2.16 | 0.919 | 6.22 | -0.553 | 5.930 | 4.642 |
| | | | | | | | | Average deviation (%) | | 1.07 |

Table 2 - Empirical coefficients for determining the lattice constants of cubic perovskites, with the expression $a_{pred} = 2\beta\ (r_B + r_X) + \gamma t - \delta$. Jiang[a] are the coefficients given in ref. [1] with incorrect β value; Jiang[b] are the coefficients with correct β value for the same work.

| | β | γ | δ | % error |
|---|---|---|---|---|
| Jiang[a] | 0.9418 | 1.4898 | 1.2062 | 0.63 |
| Jiang[b] | 0.9148 | 1.4314 | 1.0368 | 0.63 |
| This work | 0.9109 | 1.1359 | 0.7785 | 1.07 |

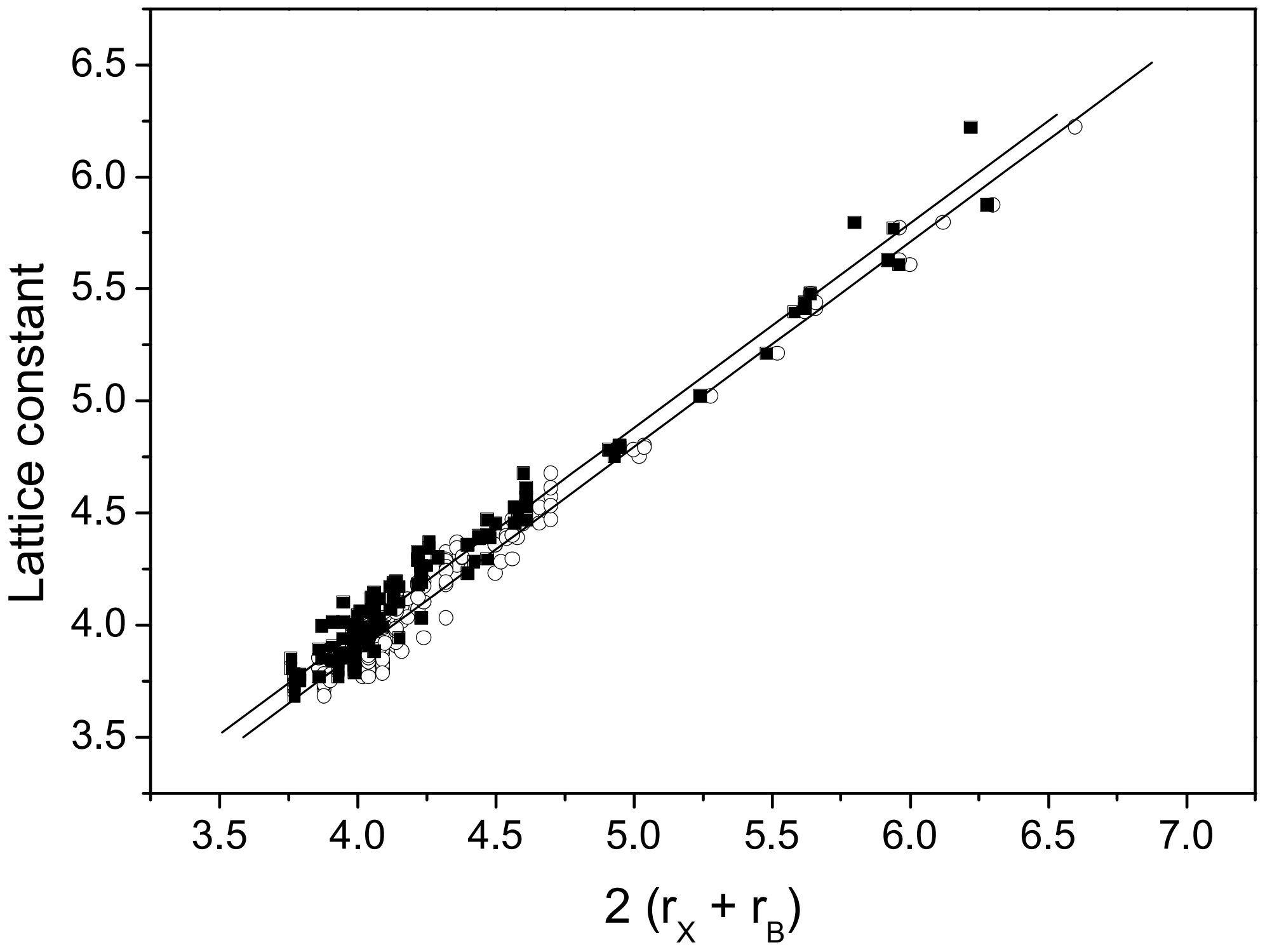

Fig. 1. Linear fits used for determining the β parameter. Here, $a = \alpha + 2\beta\ (r_B + r_X)$. Open circles for data of ref. [1]; full squares are our corrected data. Both axes in Angstroms.

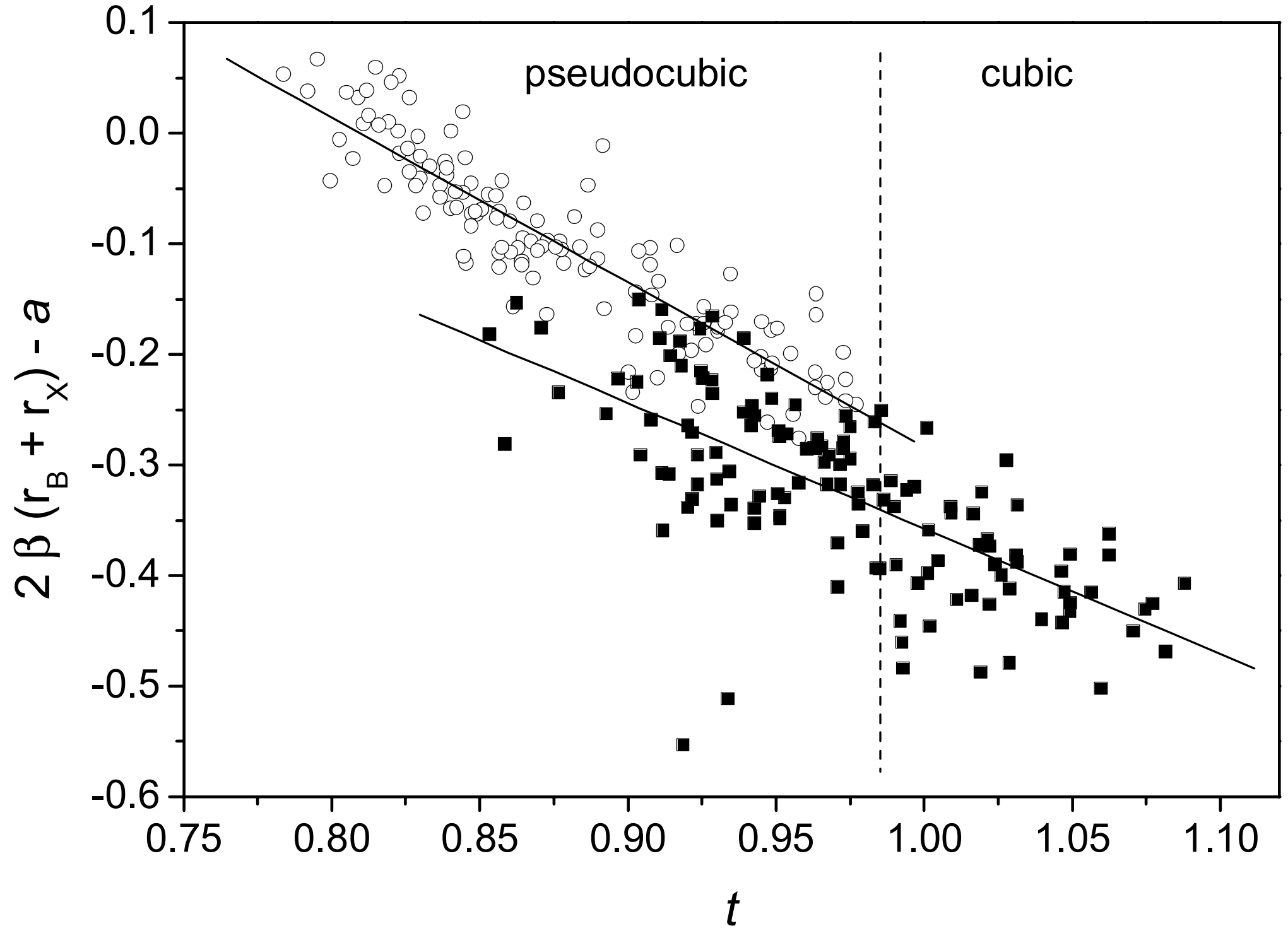

Fig. 2. Linear fits of $2\beta\ (r_B + r_X) - a = -\gamma t + \delta$, for obtaining the parameters $\gamma$ and $\delta$. Same symbols as in Fig. 1. The dashed line indicates the critical line for cubic structures. Vertical axis in Angstroms.